\documentclass[
    ,final            
  ]
  {aipproc}

\layoutstyle{6x9}

\begin{document}

\title{Radiation Damage in Polarized Ammonia Solids}

                

\author{K. Slifer\footnote{On behalf of the UVA Polarized Target Group.}}{
  address={Physcis Department, Uvinversity of Virginia, Charlottesville, VA 22903}
}

\begin{abstract}
Solid $^{15}$NH$_3$ and $^{15}$ND$_3$ provide a highly polarizable,
radiation resistant 
source of polarized protons and deuterons and have been used extensively in 
high luminosity (10$^{35}$ cm$^{-2}$s$^-1$) experiments
investigating the spin structure of the nucleon.
Over the past twenty years, the UVA polarized target group has been instrumental in
producing and polarizing much of the material used in these studies, and many practical
considerations have
been learned in this time. 
In this discussion, we analyze the polarization performance of the solid ammonia targets 
used during the recent JLab Eg4 run.
Topics include the rate of polarization decay with accumulated charge,
the annealing procedure for radiation damaged targets to recover polarization,
and the radiation induced change in optimum microwave frequency used to polarize the sample.
We also discuss the success we have had in implementing frequency modulation of the 
polarizing 
microwave frequency.
\end{abstract}

\maketitle


\section{Introduction}
The Jefferson Lab Eg4~\cite{EG4} experiment ran in 2006 with the goal of extracting 
the proton and deuteron $g_1$ structure functions, 
and the extended GDH Sum at low $Q^2$.  For this purpose we utilized the
JLab Hall B polarized target apparatus~\cite{Keith:2003ca} loaded with solid
$^{15}$NH$_3$ and $^{15}$ND$_3$.
We discuss here some phenomenological observations regarding the polarized target 
material used during the experiment. 
These observations are of a practical nature and are in general 
consistent with previous experience~\cite{McKee:2003dg,McKee:2004}.

\begin{figure}
  \includegraphics[height=.36\textheight]{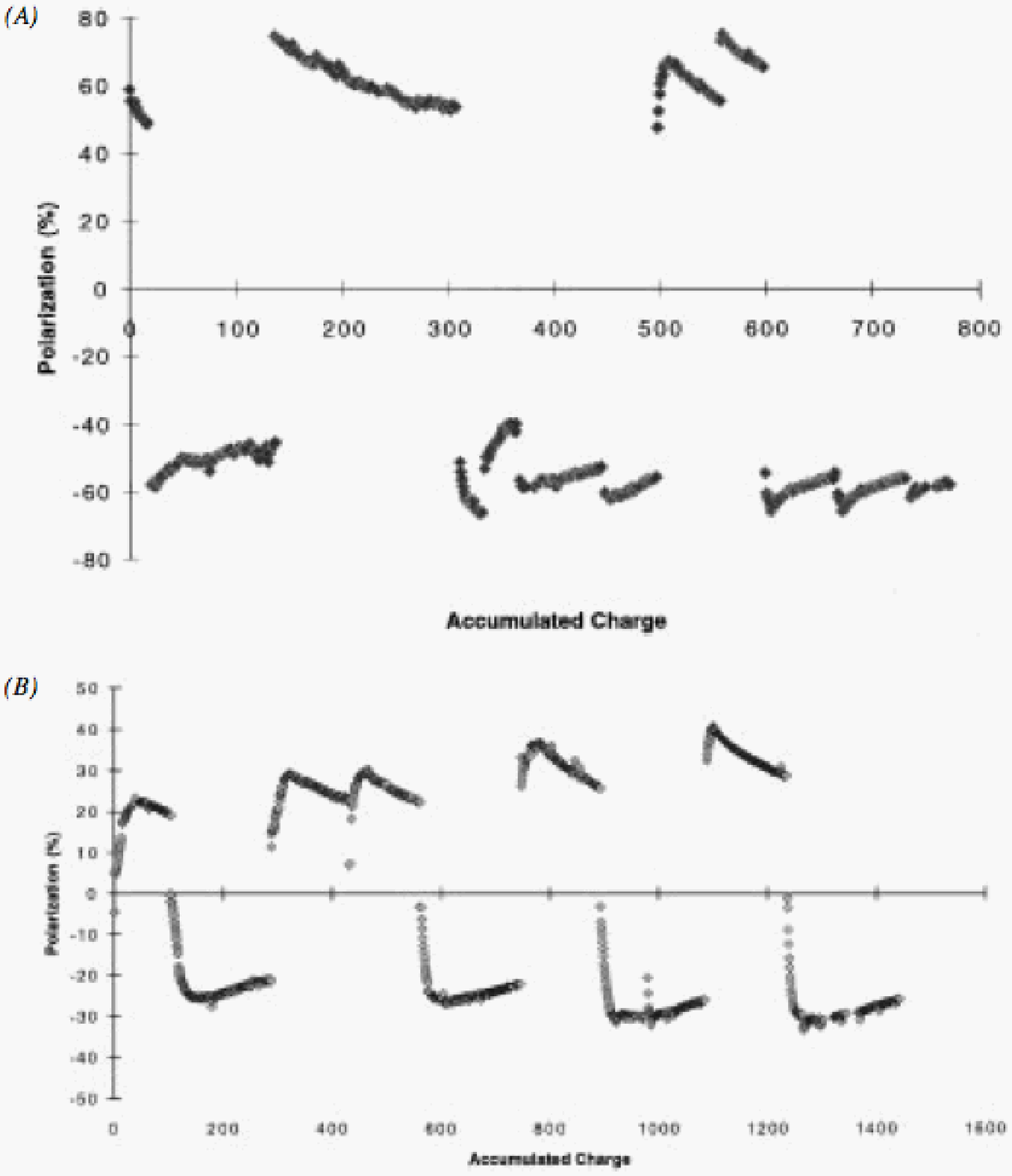}
  \includegraphics[height=.36\textheight]{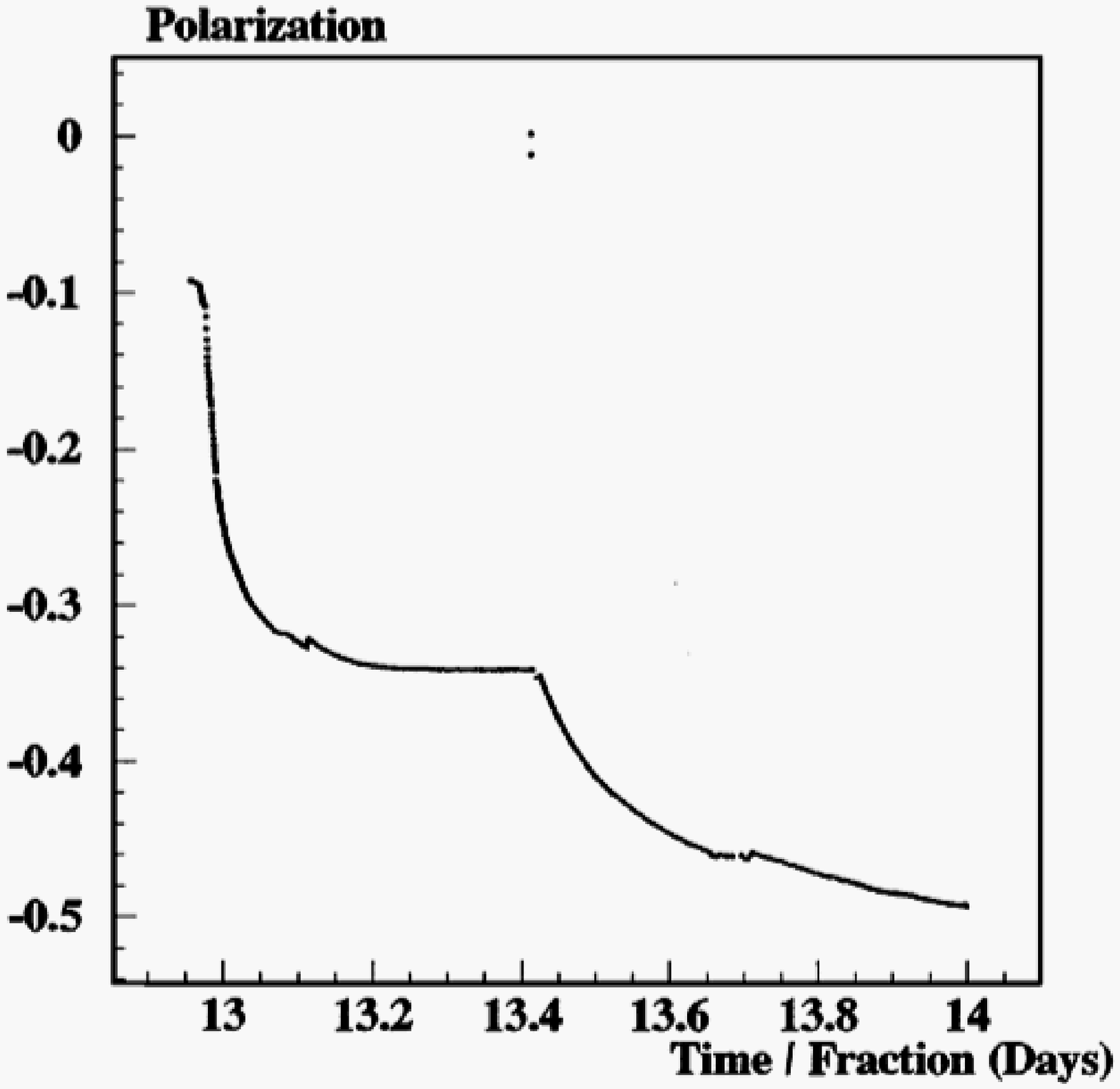} 
\caption{\label{RADDAM}
{\bf Left: }
Typical behaviour of $^{15}$NH$_3$ (top) and $^{15}$ND$_3$ (bottom) with 
accumulated dose.
{\bf Right: }
The effect of applying microwave frequency modulation to deuterated butanol during
the SMC~\cite{Adeva:1995ha,Adams:1994zd,Adams:1997hc} experiment. 
Frequency Modulation applied 
at $T\approx 13.4$. 
Both figures reproduced with
permission from ~\cite{CM}.
 }
\end{figure}

The target's 5 Tesla magnetic field is produced by a split Helmholtz pair 
superconducting magnet.  And the cooling 
power 
needed to handle the heat deposited by the JLab electron 
beam is supplied by a 1 Kelvin $^4$He evaporation refrigerator.  Approximately,
1 Watt of microwave power at 140 GHz  is used to
pump electrons during the process of Dynamic Nuclear Polarization (DNP).
Complete details of the apparatus and polarization procedure can be found in Ref.~\cite{Keith:2003ca}.

\section{Radiation Damage}


To create the paramagnetic centers that are needed for the
DNP process, raw frozen ammonia must be irradiated
with a dose of approximately 10$^{17}$ e$^-$/cm$^2$.  This {\it pre-irradiation} was performed
with the material at 87 K under liquid Argon
at a low energy electron accelerator prior to the start of the experiment.
%
With this treatment, $^{15}$NH$_3$ typically polarizes to greater than 90\% at 5 T and 1 K
on the initial spinup.

$^{15}$ND$_3$ behaves quite differently, as displayed in the lower left panel of 
Fig.~\ref{RADDAM}.
The initial deuteron polarization is typically less than 20\%,
but further {\it cold irradiation} with the target in beam at 1 K
subsequently increases the maximum $^{15}$ND$_3$ polarization to 40-45\% (5T/1K) 
after an additional cold dose of about 10$^{16}$ e$^-$/cm$^2$.

Radiation damage is the creation of unwanted radicals
in the material which do not
participate in the DNP process, but which do provide an additional 
relaxation mechanism.  As the dose accumulates, the polarization decays in
a roughly exponential fashion.  When the polarization
falls to unacceptable levels, the target is subjected to annealing.
This is the process of reducing the concentration of these unwanted radicals
in the material by warming the material to temperatures similar to the
conditions of pre-irradiation. 
Typically this process restores the maximum achievable proton polarization to the previous
best value, and  can be repeated many times before the material needs to be replaced.  
The deuteron on the other hand does not reach 
it's maximum polarization until it has been subjected to several dosing and anneal 
cycles. There is of course some risk involved with every anneal, since it is possible
to destroy the
paramagnetic centers needed for DNP if the temperature is raised too high.
During Eg4 we annealed at approximately 90K for 1 hour each time with good results.

\begin{figure}
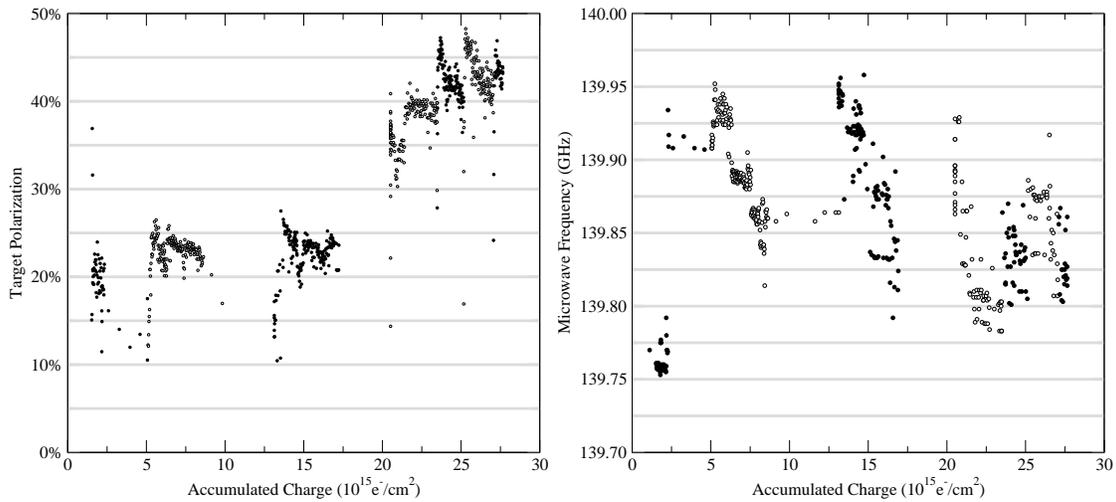

  \includegraphics[height=.3\textheight]{figs/deut_polvdose.eps}
  \includegraphics[height=.3\textheight]{figs/deut_micvdose.eps}
  \caption{\label{DMIC}
{\bf Left: } Deuteron polarization as a function of accumulated dose. 
 Note that Eg4 polarized $^{15}$ND$_3$ only in
the positive state unlike what is shown in Fig.~\ref{RADDAM}.
{\bf Right: } Eg4 Deuteron optimal microwave frequency as a function of accumulated dose. Open/closed symbols are used to indicate different anneal cycles.
}
\end{figure}

The Eg4 beam current was limited to a few nanoAmps due to the luminosity limits of the
Hall B detector package.  This is about two orders of magnitude less intense than
was used in the previous Hall C and SLAC polarized target experiments.  
To achieve 
sufficient target dosing, it was necessary to 
periodically disable the detector stack and irradiate the target with 100nA beam.
We scheduled these one hour cold dose runs as frequently
as possible during the run.  The improvement in polarization ( and figure of merit, 
which depends on the polarization squared ) is reflected in Fig.~\ref{DMIC} 
and was substantial. 

The Eg4 program was temporarily interupted by a failure of the 
JLab End Station Refrigerator (ESR)
which provides liquid cryogenic to the Halls.  During this time it was necessary to
remove the $^{15}$ND$_3$ from the target and store the material under liquid nitrogen
for several days while the ESR recovered.  This had the effect of providing an extended, 
low level anneal of the material.
When the $^{15}$ND$_3$ was replaced and subjected to a 
further cold dose, the polarization grew by 10\% as shown in Fig.~\ref{DMIC}, and
continued to grow with further dose.  
At the end of the run,
the deuteron polarization
had increased to greater than 45\%. During this time the material had absorbed less than
30 x 10$^{15}$ e$^-$/cm$^2$.  Previous experience~\cite{McKee:2003dg,McKee:2004} has shown that
ND$_3$ is extremely radiation hard, able to withstand up to
100 x 10$^{15}$ e$^-$/cm$^2$ before replacement is needed.

\section{DNP Microwave Frequency}
There are two possible DNP pumping frequencies depending on the energy
transition selected.  The initial
separation of the two frequencies is about 200 MHz, but these optimal
frequencies drift steadily apart with accumulated dose reflecting the
change in material caused by the radiation.
The approximate 100 MHz variation in optimum frequency
over each anneal cyle is shown in the right panel of Fig.~\ref{DMIC} for the $^{15}$ND$_3$
target. 
Over the first few
cycles the optimal frequency dropped sharply with dose, and returned to
it's previous initial value
after an anneal. This behaviour is not as apparent after a dose of
about $20\cdot 10^{16}$ e$^-$/cm$^2$, during the high polarization
part of Eg4.  The polarization frequency was
adjusted by hand and reflects the judgement of the target operator
as to what is optimal at that moment,  so in this regard, it is not
clear whether the increased scatter in optimal frequency during
the latter part of the experiment reflects an actual change in
material characteristics as might also be inferred from the figure.

\begin{figure}
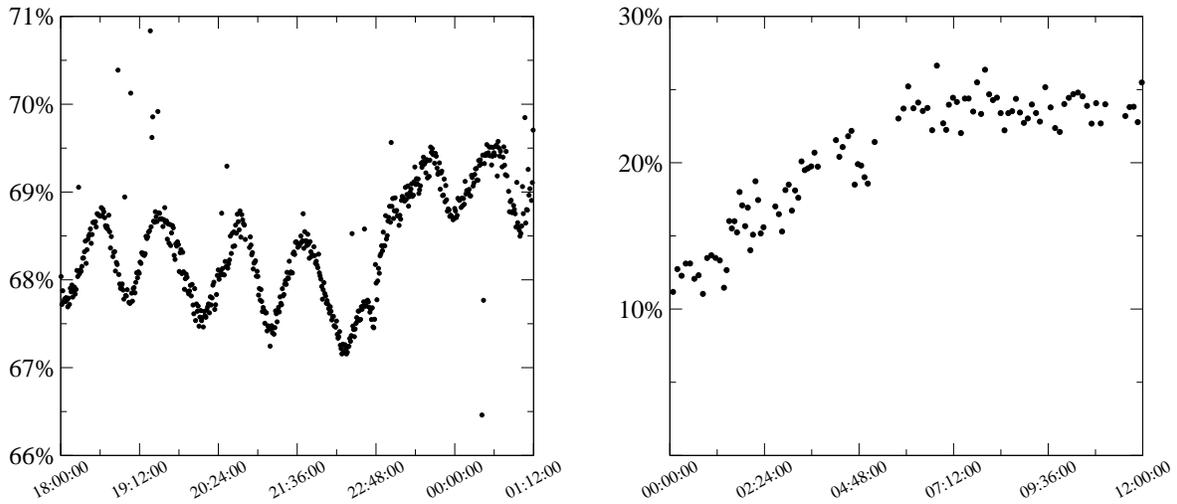

  \includegraphics[height=.3\textheight]{figs/fm_prot_may11.eps}
  \hspace{0.5cm}
  \includegraphics[height=.3\textheight]{figs/fm_deut_apr07.eps}
  \caption{\label{FM}  Polarization vs. Time showing effect of the application of Frequency Modulation to microwaves.
    {\bf Left:} Proton polarization. FM activated at approximately 22:48. Note the sinusoidal background is caused by mistuned cryogenic PID loop and is unrelated to the frequency modulation. {\bf Right:} Deuteron polarization. FM activated at approximately 04:48.}
\end{figure}

Frequency modulation of microwave frequency has been shown to have
a dramatic effect on the maximum achievable polarization.
See for example, Fig.~\ref{RADDAM} from  Ref.~\cite{CM}. 
With this in mind, we implemented a modulation of the polarizing microwave frequency
during Eg4 using a 1 kHZ 5 Volt peak-to-peak signal.  
While not as dramatic as previously published results, we did observe a few percent
improvement in polarization for both the proton and deuteron target polarization as shown
in Fig.~\ref{FM}. 

\section{Conclusion}
While $^{15}$NH$_3$ typically polarizes to greater than 90\% after an initial
warm irradiation of approximately $10^{17}$ e$^-$/cm$^2$ at 87 K,
a further series of cold irradiations and anneals are crucial to obtain large polarizations in $^{15}$ND$_3$.
Annealing for 90-100K for approximately one hour after the polarization has dropped
brings proton polarization back to previous maximum, while it typically increases
the maximum possible polarization in $^{15}$ND$_3$.

This target technology will be used again in 2008 for the SANE~\cite{SANE}, SemiSANE~\cite{SEMI}, Wide Angle Compton 
Scattering~\cite{WAC}, and $g_1^D/F_1^D$~\cite{G1D} experiments in Jefferson Lab Hall C. In addition,
the $\delta_{LT}$ experiment~\cite{DLT} which will measure the proton spin
structure function $g_2^p$ and generalized spin polarizability $\delta_{LT}$ will require
a first time installation of this target in Jefferson Lab Hall A.



%



\begin{theacknowledgments}
Thanks to D. Crabb for many informative discussions on this topic.
We also wish to acknowledge the efforts of the JLab Target group to
improve the polarized target apparatus performance, and thank them for their excellent support
during the Eg4 experiment.
This work was supported by Department of Energy contract DE-FG02-96ER40950 and
by the Institute of Nuclear and Particle Physics of the University of Virginia.
\end{theacknowledgments}

\end{document}